\documentclass[a4paper]{llncs}
\pdfoutput=1
\usepackage{afterpage}
\usepackage{booktabs}
\usepackage{color}
\usepackage{comment}
\usepackage{hyperref}
\usepackage{graphicx}
\usepackage[utf8]{inputenc}
\usepackage{listings}
\usepackage{tablefootnote}
\usepackage{wrapfig}
\usepackage[table]{xcolor}

\usepackage{chngcntr}

\hypersetup{colorlinks=true}   

\DeclareTextSymbolDefault{\dh}{T1}
\DeclareTextSymbolDefault{\DH}{T1}

\newcommand\blfootnote[1]{%
  \begingroup
  \renewcommand\thefootnote{}\footnote{#1}%
  \addtocounter{footnote}{-1}%
  \endgroup
}

\author{Finn Årup Nielsen\inst{1} \and Daniel Mietchen\inst{2} \and Egon Willighagen\inst{3}}

\institute{Cognitive Systems, DTU Compute, Technical University of Denmark, Denmark
\and  EvoMRI Communications, Jena, Germany
\and Dept of Bioinformatics - BiGCaT, NUTRIM, Maastricht University, The Netherlands}

\title{Scholia and scientometrics with Wikidata}

\begin{document}
\maketitle
\counterwithin{lstlisting}{chapter}
\renewcommand{\thelstlisting}{\arabic{lstlisting}}

\begin{abstract}
  Scholia is a tool to handle scientific bibliographic information through Wikidata. 
  The Scholia Web service creates on-the-fly scholarly profiles for researchers,
  organizations, journals, publishers, individual scholarly works, and for research topics.
  To collect the data, it queries the SPARQL-based Wikidata Query Service.
  Among several display formats available in Scholia are lists of publications for individual researchers and organizations, publications per year, employment timelines, 
  as well as co-author and topic networks and citation graphs.
  The Python package implementing the Web service is also able to
  format Wikidata bibliographic entries for use in LaTeX/BIBTeX.
\end{abstract}

\blfootnote{This article is available under the terms of the \href{https://creativecommons.org/licenses/by/4.0/}{Creative Commons Attribution 4.0 License.}}

\section{Introduction}

Wikipedia contains significant amounts of data relevant for scientometrics, and 
it has formed the basis for several scientometric studies
\cite{Q28792461,Q27966799,Q21172284,Q26857876,Q28790453,Q27467795,Q28792469}. 
Such studies can use the structured references found in Wikipedia
articles or use the intrawiki hyperlinks, e.g., 
to compare citations from Wikipedia to scholarly journals with Thomson
Reuters journal citation statistics as in \cite{Q21172284} or to rank
universities as in \cite{Q28792469}.

While many Wikipedia pages have numerous references to scientific articles, the current Wikipedias have very few entries \emph{about} specific scientific articles.
This is most evident when browsing the
\emph{Academic journal articles} category on the English Wikipedia.\footnote{\url{https://en.wikipedia.org/wiki/Category:Academic\_journal\_articles}}
Among the few items in that category are famed papers such as the 1948 physics paper
\emph{The Origin of Chemical Elements} \cite{Q21709236}~-- described in the
 English Wikipedia article \emph{Alpher–Bethe–Gamow paper}\footnote{\url{https://en.wikipedia.org/wiki/Alpher\%E2\%80\%93Bethe\%E2\%80\%93Gamow\_paper}}~-- 
as well as the 1953 article \emph{Molecular Structure of Nucleic Acids: A Structure for Deoxyribose Nucleic Acid} \cite{Q1895685} on eight Wikipedias.
Another scientific article is Hillary Putnam's \emph{Is Semantics Possible?} \cite{Q25506299}\footnote{\url{https://et.wikipedia.org/wiki/Is\_Semantics\_Possible\%3F}} from 1970 on the Estonian Wikipedia.

References in Wikipedia are often formatted in templates, and it takes some effort to
extract and match information in these fields. 
For instance, in a study of journals cited on Wikipedia, a 
database was built containing journal name variations to match the
many different variations that Wikipedia editors used when citing
scientific articles  
\cite{Q21172284}.
The use of standard identifiers~-- such as the Digital Object Identifier
(DOI)~-- in citations on Wikipedia can help to some extent to uniquely
identify works and journals.  

Several other wikis have been set up to describe scientific articles,
such as  WikiPapers, AcaWiki, Wikilit \cite{Q27778086}
and Brede Wiki \cite{Q27184103}.\footnote{\url{http://wikipapers.referata.com/}, \url{https://acawiki.org/}, \url{http://wikilit.referata.com/} and \url{http://neuro.compute.dtu.dk/wiki/}}
They are all examples of Media\-Wiki-based wikis that primarily describe scientific
articles.  
Three of them use the Semantic MediaWiki extension \cite{Q28916707}, 
while the fourth
uses MediaWiki's template functionality\footnote{\url{https://www.mediawiki.org/wiki/Help:Templates}} to structure
bibliographic information. 

Since the launch of Wikidata\footnote{\url{https://www.wikidata.org}} \cite{Q18507561}, the Wikimedia family includes a platform to better handle structured data such as
bibliographic data and to enforce input validation to a greater degree
than Wikipedia.
Wikidata data can be reified to triples \cite{Q27036482}, and 
graph-oriented databases, including SPARQL databases, can represent
Wikidata data \cite{Q27037167}.
The Wikidata Query Service (WDQS)\footnote{\url{https://query.wikidata.org}} is an
extended SPARQL endpoint that exposes the Wikidata data. 
Apart from offering a SPARQL endpoint, it also features an
editor and a variety of frontend result display options.
It may render the SPARQL query result as, e.g., bubble charts, line charts, graphs, timelines, list of images, points on a geographical map, or just provide the result as a table.
These results can also be embedded on other Web pages via an HTML iframe
element.
We note that Wikidata is open data published under \href{https://creativecommons.org/publicdomain/zero/1.0/deed.en}{CC0}, and it is available not only through the SPARQL endpoint, but also---like any other project of the Wikimedia family---through
an API and dump files.\footnote{The API is at \url{https://www.wikidata.org/w/api.php}, and the dump files are available at \url{https://www.wikidata.org/w/api.php}.}

In the following sections, we describe how Wikidata has been used for bibliographic information and present Scholia, our website built to expose such information. We furthermore show how Scholia can be used for bibliography generation and discuss limitations and advantages with Wikidata and Scholia.

\section{Bibliographic information on Wikidata}

\begin{table}[bt]
  \centering
  \begin{tabular}{lp{8cm}}
    \toprule
    Dimension & Description \\
    \midrule
    Domain & Broad coverage \\ 
    Size & $>600,000$ scientific articles \\
    Style of Metadata & Export via, e.g., Lars Willighagen's
    citation.js\tablefootnote{\url{https://github.com/larsgw/citation.js}} \\
    Persistent Inbound Links? & Yes, with the Q identifiers \\
    Persistent Outbound Links & Yes, with identifiers like DOI, PMID, PMCID, arXiv \\
    Full Text? & Via identifiers like DOI or PMCID; dedicated property for `full text URL' \\
    Access & Free access \\
    \bottomrule
  \hspace{0.2cm}
  \end{tabular}
  \caption{Summary of Wikidata as a digital library.
    This table is directly inspired by \cite[\tablename~1]{Q21092568}.}
  \label{tab:summary}
\end{table}

Wikidata editors have begun to systematically add scientific
bibliographic data to Wikidata across a broad range of scientific domains~--
see \tablename~\ref{tab:summary} for a summary of Wikidata as a digital library.
Individual researchers and scientific articles not described by their own Wikipedia article in any language are routinely added to Wikidata,
and we have so far 
experienced 
very few 
deletions of such data in reference to a notability criterion. 
The current interest in expanding bibliographic information on
Wikidata has been boosted by the WikiCite project, which aims at collecting bibliographic information in Wikidata and held its first
workshop in 2016 \cite{Q28843308}.

The bibliographic information collected on Wikidata is about
books, articles (including preprints), authors, organizations, journals, and publishers.
These items (corresponding to \emph{subject} in Semantic Web parlance) can be interlinked through Wikidata properties (corresponding to the \emph{predicate}), such as
author (P50),\footnote{The URI for Wikidata property P50 is http://www.wikidata.org/prop/direct/P50 or with the conventional prefix wdt:P50. Similarly for any other Wikidata property.}
published in (P1433), publisher (P123), series (P179), main theme (P921),
educated at (P69), employer (P108), part of (P361), sponsor (P859, can be used for funding),
cites (P2860) and several other properties.\footnote{A Wikidata table  lists properties that are commonly used in bibliographic contexts: \url{https://www.wikidata.org/wiki/Template:Bibliographical_properties} .
}

Numerous properties exist on Wikidata for deep linking to external resources,
e.g., for DOI, PMID, PMCID, arXiv, ORCID, Google Scholar, VIAF, 
Crossref funder ID, ZooBank and Twitter. 
With these many identifiers, Wikidata can act as a hub for scientometrics studies between 
resources.
If no dedicated Wikidata property exists for a resource, one of the URL
properties can work as a substitute for creating a deep link to a
resource.
For instance, P1325 (\emph{external data available at}) can 
point to raw or supplementary data associated with a paper.
We have used this scheme for scientific articles associated with
datasets stored in OpenfMRI \cite{Q21129373}, an online database with
raw brain measurements, mostly from functional magnetic resonance
imaging studies. 
Using WDQS, we query the set of OpenfMRI-linked items using the
following query: 
\begin{lstlisting}
?item wdt:P1325 ?resource .
filter strstarts(str(?resource),
                 "https://openfmri.org/dataset/")    
\end{lstlisting}
A similar scheme is used for a few of the scientific articles associated
with data in the neuroinformatics databases Neurosynth \cite{Q28916750} and NeuroVault \cite{Q25473166}.

When bibliographic items exist in Wikidata, they can be used as
references to support claims (corresponding to \emph{triplets} with extra qualifiers) in other items of Wikidata, e.g.,
a biological claim can be linked to the Wikidata item for a scientific
journal. 

By using these properties systematically according to an emerging data\break model,\footnote{\url{https://www.wikidata.org/wiki/Wikidata:WikiProject\_Source\_MetaData/Bibliographic\_metadata\_for\_scholarly\_articles\_in\_Wikidata}} editors have extended the
bibliographic information in Wikidata. Particularly instrumental in this process was a set of tools built by Magnus Manske, \emph{QuickStatements}\footnote{\url{https://tools.wmflabs.org/wikidata-todo/quick\_statements.php}} and \emph{Source MetaData},\footnote{h\url{https://tools.wmflabs.org/sourcemd/}}
including the latter's associated \emph{Resolve authors} tool.\footnote{\url{https://tools.wmflabs.org/sourcemd/new_resolve_authors.php}} 
Information can be extracted from, e.g., PubMed, PubMed Central and arXiv and added
to Wikidata. 

How complete is Wikidata in relation to scientific bibliographic
information?
Journals and universities are well represented.
For instance, 31,895 Wikidata items are linked with the identifier for the
Collections of the National Library of Medicine (P1055).
Far less covered are individual articles, individual researchers, university departments and
citations between scientific articles. 
Most of the scientific articles in Wikidata are claimed to be an 
\emph{instance of} (P31)
the Wikidata item \emph{scientific article}
(Q13442814).
With a WDQS query, we can count the number of Wikidata items linked to
\emph{scientific article}:
\begin{lstlisting}
select (count(?work) as ?count) where {
  ?work wdt:P31 wd:Q13442814 . }
\end{lstlisting}
As of 12 March 2017, the query returned the result 615,182, see also
Table~\ref{tab:summary}. 
In comparison, arXiv states having 1,240,585 e-prints
and ACM Digital Library states to have 24,110 proceedings.\footnote{As of 9 March 2017 according to \url{https://arxiv.org/} and \url{https://dl.acm.org/contents_guide.cfm}}
There were 8,617 authors associated with Wikidata items linked through
the \emph{author} property (P50) to items that are \emph{instance of}
\emph{scientific article}, and
the number of citations as counted by triples using the P2860
(\emph{cites})
property stood at 2,729,164:
\begin{lstlisting}
select (count(?citedwork) as ?count) where {
  ?work wdt:P2860 ?citedwork . }
\end{lstlisting}

The completeness can be fairly uneven. 
Articles from PLOS journals are much better represented than articles from the journals of IEEE.

The sponsor property has been used extensively for \emph{National Institute
for Occupational Safety and Health} (NIOSH) with 52,852 works linking
to the organization, 18,135 of which are \emph{instance of}
\emph{scientific articles},
but apart from NIOSH, the use of the property has been very limited for scientific
articles.\footnote{National Institute for Occupational
  Safety and Health has a Wikimedian-in-Residence program, through which James
  Hare has added many of the NIOSH works.}

\clearpage

\section{Scholia}

\begin{wrapfigure}[38]{r}{5cm}
  \centering
  \vspace{-9mm}
  \includegraphics[width=5cm]{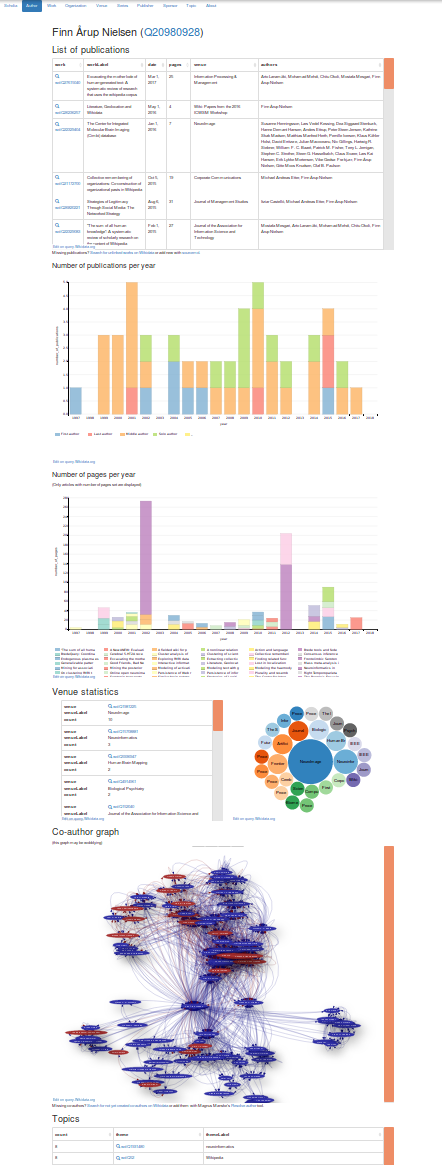}
  \caption{Overview screenshot of part of the Scholia Web page for an author:
    \url{https://tools.wmflabs.org/scholia/author/Q20980928}.
    \figurename~\ref{fig:number_of_papers_per_year} zooms in on one panel.
    }
  \label{fig:author_aspect}
\end{wrapfigure}

Scholia provides both a Python package 
and a Web service for presenting and
interacting with scientific information from Wikidata. The code is available via \url{https://github.com/fnielsen/scholia}, and a first release has been archived in Zenodo \cite{Q28936799}. 
As a Web service, its canonical site runs 
from the Wikimedia Foundation-provided service \emph{Wikimedia Tool Labs} at
\url{https://tools.wmflabs.org/scholia/}, but the Scholia package may be
downloaded and run from a local server as well. 
Scholia
uses the Flask Python Web framework \cite{Q28822647}.
The current Web service relies entirely
on Wikidata for all its presented data.
The frontend consists mostly of HTML iframe elements for embedding the 
on-the-fly-generated WDQS results and uses
many of the different output formats from this service: tables, bubble charts, bar charts, line charts, graphs and image lists.

Through a JavaScript-based query to the MediaWiki API, an excerpt from the English
Wikipedia is shown on the top of each Scholia page 
if the corresponding Wikidata item is associated with an article in the English Wikipedia.
The label for the item is fetched via Wikidata's MediaWiki API.
While some other information can be fetched this way, Scholia's many aggregation queries are better handled through SPARQL.

\begin{table}[tb]
  \centering
  \renewcommand{\arraystretch}{1.2}

  \begin{tabular}{llp{6cm}}
    \toprule
    Aspect & Examples & Example panels \\
    \midrule
    Author & Scientists & List of publications, publications per year,
                         co-authors, topics, timelines, map,
                         citations, academic tree.\\ 
    Work & Papers, books & Recent citations, citations in the work,
                           statements supported in Wikidata \\
    Organization & Universities, research groups & Affiliated authors,
                                                   co-author graph,
                                                   recent publications,
                                                   page production, co-author-normalized citations per year \\ 
    Venue & Journals, proceedings & Recent publications, topics in the
    publications, author images, prolific authors, most cited works,
    most cited authors, most cited venues \\
    Series & Proceedings series & Items (venues) in the series,
    published works from venues in the series \\
    Publisher & Commercial publisher & Journals and other publications
    published, associated editors, most cited papers, number of
    citations as a function of number of published works \\
    Sponsor & Foundation & List of publications funded, sponsored
                           authors, co-sponsors\\
    Topic & Keywords & Recent publication on the topic, co-occurring
    topics \\
    \bottomrule
    \hspace{0.2cm}
  \end{tabular}
  \caption{Aspects in Scholia: Each Wikidata item can be viewed in one
    or more aspects. Each aspect displays multiple ``panels'', which
    may be, e.g., a table of publications or a bar chart of citations per year.}
  \label{tab:aspects}
\end{table}

Scholia uses the Wikidata item identifier as its identifier rather than
author name, journal titles, etc.
A search field on the front page provides a Scholia user with the ability to
search for a name to retrieve the relevant Wikidata identifier.
To display items, Scholia sets up a number of what we call ``aspects''. 
The currently implemented aspects are author, work, organization, venue,
series, publisher, sponsor and topic, see \tablename~\ref{tab:aspects}.
The present selection was motivated by the possibilities inherent in the Wikidata items and properties.
We plan to extend this to further aspects, e.g., award or determination method.
A URL scheme distinguishes the different aspects,
so the URL path /scholia/author/Q6365492 will show the author aspect of the
statistician Kanti V. Mardia, while /scholia/topic/Q6365492 will show
the topic aspect of the person, e.g., articles about Mardia.
Likewise, universities can be viewed, for instance, as organizations or as sponsors.
Indeed, any Wikidata item can be viewed in any Scholia aspect,
but Scholia can show no data if the user selects a ``wrong'' aspect, i.e. one for which no relevant data is available in Wikidata.

For each aspect, we make multiple WDQS queries based on the
Wikidata item for which the results in the panels are displayed, --- technically in embedded iframes.
For the author aspect, Scholia queries WDQS for the list of
publications, showing the result in a table, displaying a bar chart of
the number of publications per year, number of pages per year, venue
statistics, co-author graph, topics of the published works (based on the ``main theme'' property), associated
images, education and employment history as timelines, academic tree, map with
locations associated with the author, and citation statistics~-- see
\figurename~\ref{fig:author_aspect} for an example of part of an author
aspect page.
The citation statistics displays the most cited work, citations by year
and citing authors.
For the academic tree and the citation graph, we make use of Blazegraph's graph analytics  RDF GAS API\footnote{\url{https://wiki.blazegraph.com/wiki/index.php/RDF_GAS_API}}
that is available in WDQS.
The embedded WDQS results link back to WDQS where a user can modify the query. 
The interactive editor of WDQS allows users not familiar with SPARQL to make simple modifications without directly editing the SPARQL code.

\begin{figure}[tb]
  \centering
  \includegraphics[width=\textwidth]{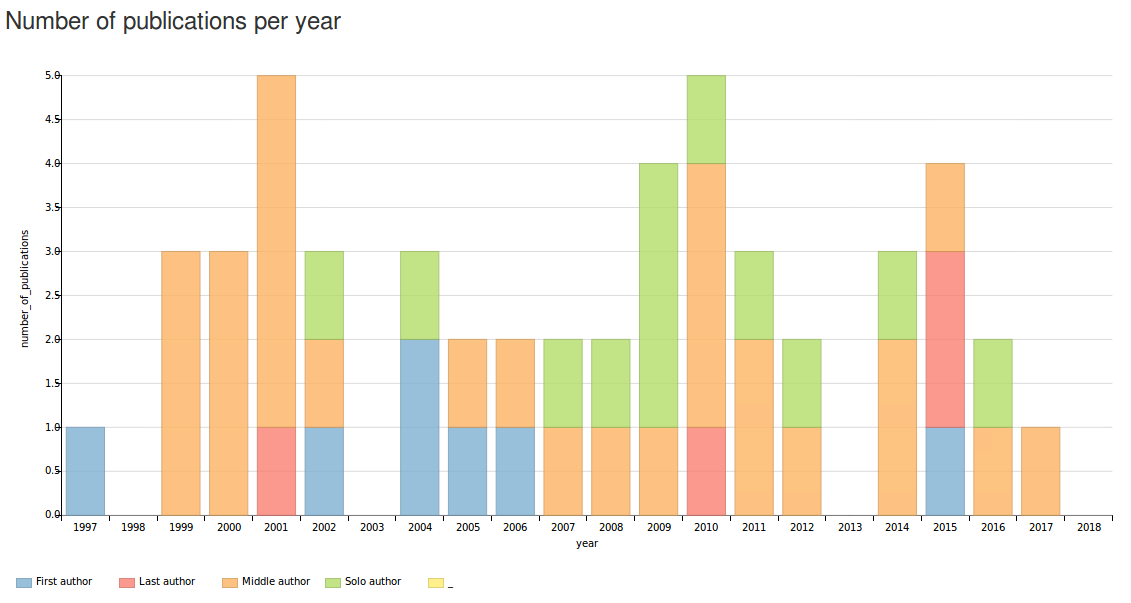}
  \caption{Screenshot of Scholia Web page with the number of papers
    published per year for Finn Årup Nielsen: 
    \url{https://tools.wmflabs.org/scholia/author/Q20980928}.
    Inspired by LEGOLAS. Colors indicate author
    role: first, middle, last or solo author.}
  \label{fig:number_of_papers_per_year}
\end{figure}

Related to their work on quantifying conceptual novelty in the
biomedical literature \cite{Q28792033}, Shubhanshu Mishra and Vetle
Torvik have set up a website profiling authors in PubMed
datasets: LEGOLAS.\footnote{\url{http://abel.lis.illinois.edu/legolas/}} 
Among other information, the website shows  
the number of articles per year, the number of citations per year,
the number of self-citations per year, unique collaborations per year and
NIH grants per year as bar charts that are color-coded according to, e.g., author
role (first, solo, middle or last author).
Scholia uses WDQS for LEGOLAS-like plots.
Figure~\ref{fig:number_of_papers_per_year} displays one such example
for the number of published items as a function of year of
publication on an author aspect page, where the components of the bars
are color-coded according to author role.

\begin{figure}[tb]
  \centering
  \includegraphics[width=\textwidth]{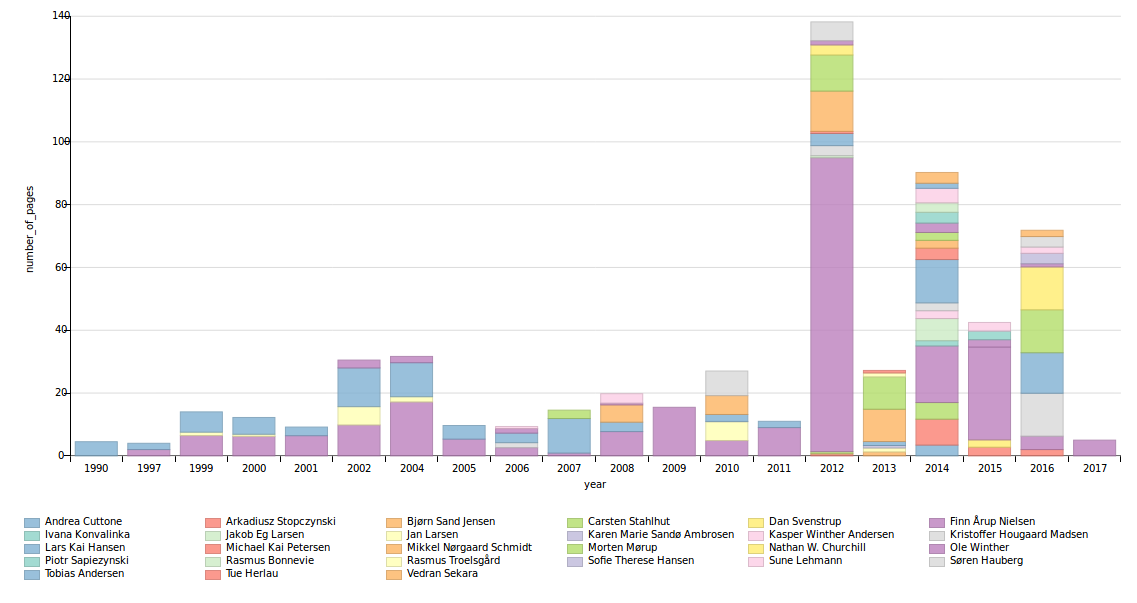}
  \caption{Scholia screenshot with page production for a research section (Cognitive Systems
    at the Technical University of Denmark), where the number of
    pages per paper has been normalized by the number of authors.
    The bars are color-coded according to author.
    The plot is heavily biased, as only a very limited subset of papers
    from the section is available in Wikidata, and the property for the  
    number of pages is set for only a subset of
    these papers.
    From \url{https://tools.wmflabs.org/scholia/organization/Q24283660}.}
  \label{fig:page_production}
\end{figure}

For the organization aspect, Scholia uses the employer and
affiliated Wikidata properties to identify associated authors, and combines this with the author query for works.
Scholia formulates SPARQL queries with property paths to identify suborganizations of the 
queried organization, such that authors affiliated with a suborganization 
are associated with the queried organization.
Figure~\ref{fig:page_production} shows a corresponding bar chart, again inspired by
the LEGOLAS style.
Here, the Cognitive Systems
section at the Technical University of Denmark is displayed with the
organization aspect.
It combines work and author data.
The bar chart uses the P1104 (number of pages) Wikidata
property together with a normalization based on the number of authors
on each of the work items.
The bars are color-coded according to individual authors associated with the
organization. 
In this case, the plot is heavily biased, as only a very limited
subset of publications from the organization is currently present in 
Wikidata, and even the available publications may not have the
P1104 property set.
Other panels shown in the organization aspect are a co-author
graph, a list of recent publications formatted in a table, a bubble
chart with most cited papers with affiliated first author and a bar
chart with co-author-normalized citations per year.
This last panel counts the number of citations to each work
and divides it by the number of authors on the cited work, then groups
the publications according to year and color-codes the bars according to
author. 

\begin{figure}[tb]
  \centering
  \includegraphics[width=\textwidth]{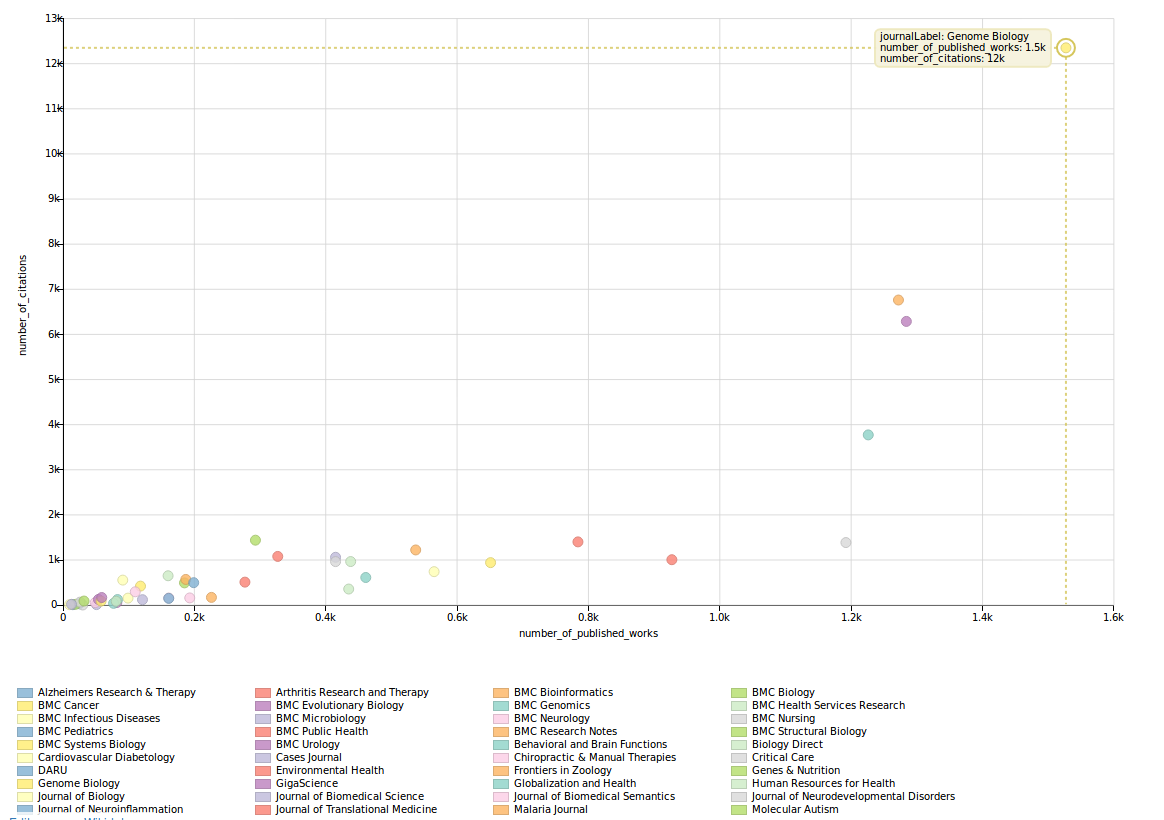}
  \caption{Screenshot from Scholia's publisher aspect with
    number of publications versus number of citations for works
    published by BioMed Central.
    The upper right point with many citations and many published works
    is the journal \emph{Genome Biology}.
    From \url{https://tools.wmflabs.org/scholia/publisher/Q463494}.} 
  \label{fig:bmc_counts}
\end{figure}

For the publisher aspect, Scholia queries all items where the P123
property (publisher) has been set.
With these items at hand, Scholia can create lists of venues (journals
or proceedings) ordered according to the number of works (papers)
published in each of them, as well as lists of works ordered according
to citations.
\figurename~\ref{fig:bmc_counts} shows an example of a panel on the
publisher aspect page with a scatter plot detailing journals from 
\emph{BioMed Central}.
The position of each journal in the plot reveals impact factor-like information.

\begin{lstlisting}[float,caption={SPARQL query on the work aspect page for claims supported by a work, --- in this case Q22253877 \cite{Q22253877}.},label=lst:supported,language=sql,frame=tb,xleftmargin=0mm]
SELECT distinct ?item ?itemLabel ?property ?propertyLabel
       ?value ?valueLabel WHERE {
  ?item ?p ?statement .
  ?property wikibase:claim ?p . 
  ?statement ?a ?value .
  ?item ?b ?value . 
  ?statement prov:wasDerivedFrom/
    <http://www.wikidata.org/prop/reference/P248>
    wd:Q22253877 .
  SERVICE wikibase:label {
    bd:serviceParam wikibase:language "en" }
} ORDER BY ?itemLabel
\end{lstlisting}

\begin{figure}[tb]
  \centering
  \includegraphics[width=0.8\textwidth]{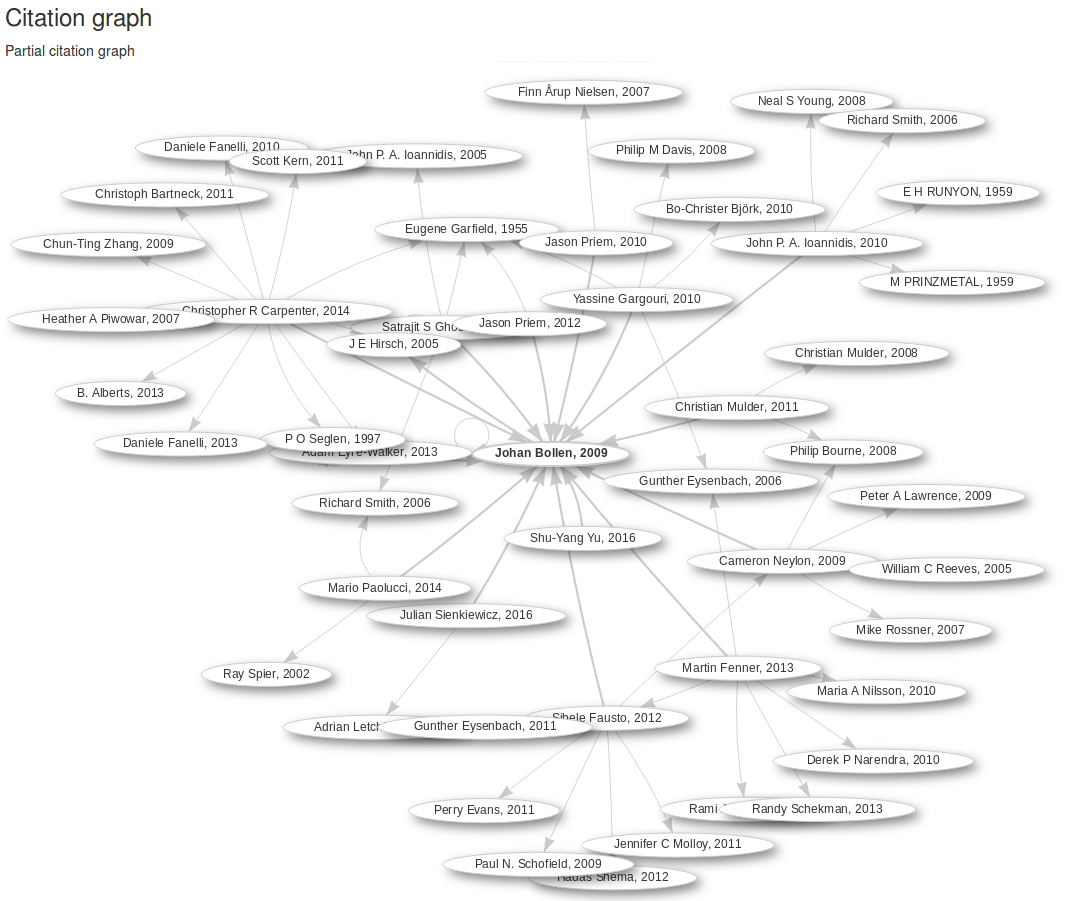}
  \caption{Screenshot of part of a Scholia Web page at  \url{https://tools.wmflabs.org/scholia/work/Q21143764} with the partial citation graph panel of the work aspect for Johan Bollen's article from 2009 \cite{Q21143764}.} 
  \label{fig:citationgraph}
\end{figure}

For the work aspect, Scholia lists citations and produces a partial citation graph.
\figurename~\ref{fig:citationgraph} shows a screenshot of the citation graph panel from the work aspect for a specific article \cite{Q21143764}. 
For this aspect, we also formulate a special query to return a table with a list of Wikidata items where the given work is used as a source for claims. 
An example query for a specific work is shown with Listing~\ref{lst:supported}.
From the query results, it can be seen, for instance, that the article \emph{A novel family of mammalian taste receptors} \cite{Q22253877} supports 
a claim about \emph{Taste 2 receptor member 16} (Q7669366) being present in the cell component (P681) \emph{integral component of membrane} (Q14327652).
For the topic aspect, Scholia uses a property path SPARQL query to identify subtopics. 
For a given item where the aspect is not known in advance, Scholia tries
to guess the relevant aspect by looking at the \emph{instance of}
property. 
The Scholia Web service uses that guess for redirecting, so /scholia/Q8219
will redirect to /scholia/author/Q8219, the author aspect for the 
psychologist Uta Frith. This is achieved by first making a server site query to establish
that Uta Frith is a human and then using that information to choose the
author aspect as the most relevant aspect to show information about
Uta Frith.

A few redirects for external identifiers are also implemented.
For instance, with Uta Frith's Twitter 
name `utafrith', /scholia/twitter/utafrith will redirect to
/scholia/Q8219, which in turn will redirect to
/scholia/author/Q8219. 
Scholia implements similar functionality for DOI, ORCID and GitHub user
identifier. 

\section{Using Wikidata as a bibliographic resource}
As a command-line tool, Scholia provides a prototype tool that uses
Wikidata and its bibliographic data in a \LaTeX\ and \textsc{Bib}\TeX\
environment.
The current implementation looks up citations in the latex-generated
\texttt{.aux} file and queries Wikidata's MediaWiki API to get
cited Wikidata items.
The retrieved items are formatted and written to a \texttt{.bib} that
bibtex can use to format the 
bibliographic items for inclusion in the \LaTeX\ document. 
The workflow for a \LaTeX\ document with the filename
example.tex is 
\begin{lstlisting}[language=sh]
latex example
python -m scholia.tex write-bib-from-aux example.aux
bibtex example
latex example
latex example
\end{lstlisting}
Here, the example document could read
\begin{lstlisting}
\documentclass{article}
\usepackage[utf8]{inputenc}
\begin{document}
\cite{Q18507561}
\bibliographystyle{plain}
\bibliography{example}
\end{document}
\end{lstlisting}

In this case, the \verb!\cite! command cites Q18507561
(\emph{Wikidata: a free collaborative knowledgebase} \cite{Q18507561}).
A DOI can also be used in the \verb!\cite! command: 
instead of writing \verb!\cite{Q18507561}!, one may
write \verb!\cite{10.1145/2629489}! to get the same citation.
Scholia matches on the ``\texttt{10.}'' DOI prefix and makes a SPARQL
query to get the relevant Wikidata item.

The scheme presented above can take advantage of the many available
style files of \textsc{Bib}\TeX\ to format the bibliographic items in
the various ways requested by publishers.
We have used Scholia for reference management in this paper.

\section{Discussion}
WDQS and Scholia can provide many different scientometrics views of
the data available in Wikidata.
The bibliographic data in Wikidata are still quite limited,
but the number of scientometrically 
relevant items will likely continue to grow considerably in the coming months and years. 

The continued growth of science data on Wikidata can have negative
impact on Scholia, making the on-the-fly queries too resource demanding. 
In the current version, there are already a few queries that run into WDQS's time out,
e.g., it happens for the view of co-author-normalized
citations per year for Harvard University.  
If this becomes a general problem, we will need to redefine the queries.
Indeed, the WDQS time out will be a general problem if we want to perform large-scale scientometrics studies. An alternative to using live queries would be using dumps, which are available in several formats on a weekly basis, with daily increments in between.\footnote{\url{https://www.wikidata.org/wiki/Wikidata:Database\_download}}
The problem is not a limitation of SPARQL, but a limitation set by the server resources.
Some queries may be optimized, especially around the item labeling. 

Working with Scholia has made us aware of several issues.
Some of these are minor limitations in the Wikidata and WDQS
systems.  
The Wikidata label length is limited to 250 characters, whereas the `monolingual text' datatype used for the `title' property (P1476) is limited to 400 characters.
There are scholarly articles with titles longer than those limits.

Wikidata fields cannot directly handle subscripts and superscripts,
which commonly appear in titles of articles about chemical
compounds, elementary particles or mathematical formulas. 
Other formatting in titles cannot directly be handled in Wikidata's
title property,\footnote{By way of an example, consider the article ``A
  library of 7TM receptor C-terminal tails. Interactions with the
  proposed post-endocytic sorting proteins ERM-binding phosphoprotein
  50 (EBP50), N-ethylmaleimide-sensitive  
  factor (NSF), sorting nexin 1 (SNX1), and G protein-coupled
  receptor-associated sorting protein (GASP)'', an article with the
  title ``Cerebral 5-HT$_{\mbox{\small 2A}}$ receptor binding is increased in
  patients with Tourette's syndrome'', where ``2A'' is subscripted and
  ``User's Guide to the \texttt{amsrefs} Package'',  where the
  ``amsrefs'' is set in monospaced font.} 
  and recording a date such as ``Summer 2011'' is difficult.

Title and names of items can change. 
Authors can change names, e.g.\ due to marriage, and journals can
change titles, e.g.\ due to a change of scope or transfer of ownership.
For instance, the \emph{Journal of the Association for Information Science and Technology} has changed names several times over the years.\footnote{\url{http://onlinelibrary.wiley.com/journal/10.1002/(ISSN)2330-1643/issues} records these former titles: \emph{Journal of the American Society for Information Science and Technology}, \emph{Journal of the American Society for Information Science}, and \emph{American Documentation}.}
Wikidata can handle multiple titles in a single Wikidata item and with
qualifiers describe the dates of changes in title. 
For scientometrics, this ability is an advantage in principle, but multiple titles
can make it cumbersome to handle when Wikidata is used as a
bibliographic resource in document preparation, particularly for articles published near the time when the journal changed its name. One way to alleviate this problem would be to split the journal's Wikidata item into several, but this is not current practice.

In Wikidata, papers are usually not described to be affiliated with
organizations. 
Scholia's ability to make statistics on scientific articles
published by organization is facilitated by the fact that items about scientific articles can link to items about
authors, which can link to items about organizations. 
It is possible to link scientific articles to organization by using
Wikidata qualifiers in connection with the author property.
However, this scheme is currently in limited use. 

This scarcity of direct affiliation annotation on Wikidata items about articles means that scientometrics on the organizational level are unlikely to be precise at present.
In the current version, Scholia even ignores any temporal qualifier for the affiliation and employer property, meaning that a researcher moving between 
several organization gets his/her articles counted under multiple organizations.

Data modeling on Wikidata gives rise to reflections on what precisely
a ``publisher'' and a ``work'' is.
A user can set the \emph{publisher} Wikidata property of a work 
to a corporate group, a subsidiary or possibly an imprint.
For instance, how should we handle \emph{Springer Nature},
\emph{BioMed Central} and \emph{Humana Press}?

\emph{Functional Requirements for Bibliographic Records} (FRBR)
\cite{Q28843737} suggests a scheme for works, expressions, manifestations and ``items''.
In Wikipedia, most items are described on the work level as opposed to the manifestation level (e.g., book edition), 
while citations should usually go to the manifestation level.
How should one deal with scientific articles that have slightly different ``manifestations'', such as preprint, electronic journal edition, paper edition and postprint, or editorials that were co-published in multiple journals with identical texts? 
An electronic and a paper edition may differ in their dates of publication, but otherwise 
have the same bibliographic data, while a preprint and its journal edition usually have different identifiers and may also differ in content.  
From a scientometrics point of view, these difference in manifestation may not matter in some cases, but could be the focus of others.
Splitting a scientific article as a work (in the FRBR sense) over multiple Wikidata items
seems only to complicate matters.

\begin{table}[h]
{\small
  	\begin{tabular}{p{4cm}cp{7.4cm}}
		\toprule
        Feature &  & Description \\
        \midrule
        Business model & Y & Community donations  and funding from foundations to Wikimedia Foundation and affiliated chapters \\
        Portrait picture & Y & The P18 property can record Wikimedia Commons images related to a researcher \\
        Alternative names & Y & Aliases for all items, not just researchers \\
        IDs / profiles in other systems & Y & Numerous links to external identifiers: ORCID, Scopus, Google Scholar, etc. \\
        
        Papers and similar & Y & Papers and books are individual Wikidata items \\
        Uncommon research products & Y & For instance, software can be associated with a developer \\
        Grants, third party funding & (N) & Currently no property for grant holders and probably no individual grants in Wikidata. The sponsor property can be used to indicate the funding of a paper \\

       Current institution & Y & Affiliation and employer can be recorded in Wikidata \\
       Former employers, education & Y & Education, academic degree can be specified, and former employers can be set by way of qualifiers \\
      	
        Self-assigned keywords & (Y) & The main theme of a work can be specified, interests or field of work can be set for a person. The values
        must be items in Wikidata. Users can create items. \\
        Concepts from controlled vocabulary & Y & See above \\
        
        Social graph of followers/friends & N & There are no user accounts on the current version of Scholia. \\
        Social graph of coauthors & Y & \\
        
        Citation/attention metadata from platform itself & Y & Citations between scientific articles are recorded with a property that can be used to count citations.
        Citation/reference between Wikidata items.
        	\\
        Citation/attention metadata from other source & (N) & Deep links to other citation resources like Google Scholar and Scopus.
        \\
        Comprehensive search to match/include papers & (N) & Several tools liks Magnus Manske's \emph{Source MetaData} that look up bibliographic metadata based on DOI, PMID or PMCID\\
        Forums, Q\&A etc. & N & \\
        Deposit own papers & (Y) & Appropriately licensed papers can be uploaded to Wikimedia Commons or Wikisource\\
        Research administration tools & N & \\
              
       	Reuse of data from outside of the service & Y & API, WDQS, XML dump, third-party services \\
       
        \bottomrule
	\end{tabular}
    }
    \caption{Overview of Wikidata and Scholia features in terms of a scholarly profile.  Directly inspired by a blog post by Lambert Heller (see text).}
   \label{tab:profile}
\end{table}

\afterpage{\clearpage}

The initial idea for Scholia was to create a researcher profile based on Wikidata data
with list of publications, picture and CV-like information.
The inspiration came from a blog post by Lambert Heller: \emph{What will the scholarly profile page of the future look like? Provision of metadata is enabling experimentation.}\footnote{\url{http://blogs.lse.ac.uk/impactofsocialsciences/2015/07/16/scholarly-profile-of-the-future/}}
In this blog post, he discussed the different features of several scholarly Web services: ORCID, ResearchGate, Mendeley, Pure, VIVO, Google Scholar and ImpactStory.
In \tablename~\ref{tab:profile}, we have set up a table listing Heller's features for the Wikidata--Scholia combination.
Wikidata--Scholia performs well in most aspects, but in the current version, Scholia has no backend
for storing user data, and 
user features such as forum, Q\&A and followers are not available. 

Beyond the features listed by Heller, which features set Wikidata--Scholia apart from 
other scholarly Web services?
The collaborative nature of Wikidata means that Wikidata users can create items for authors that do not have an account on Wikidata.
In most other systems, the researcher as a user of the system has control over 
his/her scholarly profile and other researchers/users cannot make amendment or corrections. Likewise, when one user changes an existing item, this change will be reflected in subsequent live queries of that item, and it may still be in future dumps if not reverted or otherwise modified before the dump creation.

With WDQS queries, Scholia can combine data from different types of items in Wikidata
in a way that is not usually possible with other scholarly profile Web services.
For instance, Scholia generates lists of publications for an organization by
combining items for works and authors and can show co-author graphs restricted by affiliation.
Similarly, the co-author graph can be restricted to authors publishing works annotated with a specific main theme.
Authors are typically annotated with gender in Wikidata, so Scholia can show
gender color-coding of co-author graphs.
On the topic aspect page, the Scholia panel that shows the most cited works that are cited from works around the topic
can point to an important paper for a topic~-- even if the paper has not been annotated with 
the topic~-- by combining the citations data and topic annotation.
References for claims are an important part of Wikidata and also singles Wikidata out 
among other scholarly profile Web service, and it acts as an extra scientometrics dimension.
The current version of Scholia has only a single panel where the query uses references:
the ``Supports the following statement(s)'' on the work aspect page, 
but it is possible to extend the use of this scientometrics dimension.

\section{Acknowledgements}

This work was supported by Innovationsfonden through the DABAI project.
The work on Scholia was spawned by the WikiCite project \cite{Q28843308}. 
We would like to thank the organizers of the workshop, particular
Dario Taraborelli.
Finn Årup Nielsen's participation in the workshop was sponsored by an
award from the Reinholdt W.\ Jorck og Hustrus Fund.
We would also like to thank Magnus Manske and James Hare for considerable work 
with Wikidata tools and data in the context of WikiCite.


\end{document}